\documentclass[12pt,thmsa]{article}
\usepackage{amssymb}
\usepackage{amsmath}
\usepackage{sw20lart}

\input{tcilatex}

\begin{document}

\author{F. Iddir$\thanks{
E-mail: \emph{faridaghis@hotmail.com}}${\large \ \ }and {\large \ }L. Semlala%
$\thanks{
E-mail: \emph{l\_semlala@yahoo.fr}}$ \and $Laboratoire${\normalsize \ }$de$%
{\normalsize \ }$Physique${\normalsize \ }$Th\acute{e}orique,${\normalsize \ 
}$Univ.${\normalsize \ }$d^{\prime }Oran${\normalsize \ }$Es${\normalsize -}$%
S\acute{e}nia${\normalsize \ }$31100$ \and $ALGERIA$}
\title{{\LARGE Hybrid States from Constituent Glue Model}}
\maketitle
\date{}

\begin{abstract}
The hybrid meson is one of the most interesting new hadron specie beyond the
naive quark model. It acquire a great attention both from the theoretical
and experimental efforts. Many good candidates have been claimed to be
observed, but there is no absolute confirmation about existence of hybrid
mesons.

In the present work we propose new calculations of the masses and decay
widths of the hybrid mesons in the context of constituent gluon model.
\end{abstract}

\section{Introduction}

There is an intensive experimental activity attempting to detect new hadrons
beyond the quark model which are predicted by QCD: glueballs, hybrids,
diquonia, pentaquarks, ...

Hybrid mesons (quark-antiquark-gluon) can have $J^{PC}$ quantum numbers
which are not allowed by the naive quark model, like $0^{--}$, $0^{+-}$, $%
1^{-+}$, $2^{+-}$,...then they can't mix with the standard mesons and hence
can facilitate their observation.

These \textquotedblright exotic\textquotedblright\ objects are the most
promising new species of hadrons allowed by QCD and subject of lot of works
both in the theoretical and experimental levels.

From experimental efforts at IHEP$^{\left[ 1\right] }$, KEK$^{\left[ 2\right]
}$, CERN$^{\left[ 3\right] }$ and BNL$^{\left[ 4\right] }$ several $J^{PC}$=$%
1^{-+}$exotic resonances have been claimed to be identified, especially $\pi
_{1}(1600)$ and $\pi _{1}(1400)$ have received great interest, but some
doubts are raised about the later one$^{\left[ 5-6\right] }$.

In the charm sector, the most interesting new states observed are the
X(3872) and the Y(4260)$^{\left[ 7\right] .}$.

This work follows and completes several papers already devoted to this
subject$^{[9-10]},$ new spin dependent estimations of the masses of hybrids
and its\ consequences to the decay widths are proposed.

In fact the hybrid mesons are studied from different models: lattice QCD$^{%
\left[ 11\right] }$, flux tube model$^{\left[ 12\right] }$, bag model$^{%
\left[ 13\right] }$, QCD sum rules$^{\left[ 14\right] }$ and constituent
gluon model$^{\left[ 7-10,15-16\right] }$. Some of them can perform both
estimations of mass and decay widths. The nature of gluonic field inside
hybrid is not yet be clear because the gluon plays a double role: it
propagates the interaction between color sources and being itself colored it
undergoes the interaction. Whereas, LQCD and Sum rules QCD make no
assumptions about it, two important hypothesis can be retained from
literature. The first one consider gluonic $\deg $rees of freedom as
\textquotedblright excitations\textquotedblright\ of the \textquotedblright
flux tube\textquotedblright\ between quark and antiquark, which leads to the
linear potential, that is familiar from quark model (flux-tube model).

The second issue, which are supported by the present work, assumes that
hybrid is a bound state of quark-antiquark and a constituent glue which
interact through a phenomenological potential, precisely Coulomb plus linear
potential supplemented by spin-spin , spin-orbit and tensor correction
terms. The use of relativistic kinetics is appropriate for the study of the
light flavor systems. The wave functions obtained are used for the
calculations of the strong decay widths.

\section{Hybrid states and binding Hamiltonian}

For the representation of hybrid states we will use the notations:

\emph{l}$_{\text{g }}$\emph{\ \ }: is the relative orbital momentum of the
gluon in the \emph{q\={q}} center of mass;

\emph{l}$_{\text{\textit{q\={q}}}}$\emph{\ \ }: is the relative orbital
momentum between \emph{q} and \emph{\={q}};

\emph{S}$_{\text{\textit{q\={q}}}}$\emph{\ }: is the total quarks spin;

\emph{j}$_{\text{\textit{g }}}$\emph{\ \ }: is the total gluon angular
momentum;

\emph{L \ \ }: \textit{l}$_{\text{\textit{q\={q}}}}$ $\oplus $ \textit{j}$_{%
\text{\textit{g}}}.$

Considering the gluon moving in the framework of the $q\overline{q}$ pair,
the Parity of the hybrid will be:

\begin{center}
\begin{eqnarray}
P &=&P\left( q\overline{q}\right) \cdot P\left( g\right) \cdot P\left(
relative\right)  \notag \\
&=&\left( -\right) ^{l_{q\overline{q}}+1}\cdot \left( -1\right) \cdot \left(
-\right) ^{l_{g}};
\end{eqnarray}
\end{center}

$\left( -1\right) $ being the intrinsic parity of the gluon. Then the Parity
of hybrid meson will be:

\begin{center}
\begin{equation}
P=\left( -\right) ^{l_{q\bar{q}}+l_{g}}.
\end{equation}
\end{center}

The Charge Conjugation is given by: 
\begin{equation}
C=\left( -\right) ^{l_{q\bar{q}}+S_{q\bar{q}}+1}.
\end{equation}

\underline{\emph{The} \emph{QE and GE hybrids}}

$S_{q\bar{q}}$ can takes values 0 or 1; \textit{P} and \textit{C} impose
Parity restrictions to $l_{q\bar{q}},\ $and $l_{g}$ (Table 1).

For lower values of orbital excitations ( $l_{q\bar{q}}$ and $l_{g}\leqslant
1$) and

\textit{P = -1}, hybrid states can be built by two modes; \textit{l}$_{\text{%
\textit{q\={q}}}}=1$ and \textit{l}$_{\text{\textit{g}}}=0,$ which we shall
refer as \emph{the quark-excited hybrid (}QE\emph{)}, and \textit{l}$_{\text{%
\textit{q\={q}}}}=0$ and \textit{l}$_{\text{\textit{g}}}=1,$ which we shall
refer as \emph{the gluon-excited hybrid (}GE\emph{).}These particular modes%
\emph{\ }play an important role in the decay selection rules to be mentioned
later. The possible mixed QE-GE state is taked into account in this paper.

\underline{\emph{The wavefunctions}}

The total wave function can be written as:

\begin{equation}
\Psi _{JM}^{PC}=\left( \left( \left( \mathbf{e}_{\mu _{g}}\otimes \psi
_{l_{g}}^{m_{g}}\right) _{j_{g}M_{g}}\otimes \psi _{l_{q\overline{q}}}^{m_{q%
\overline{q}}}\right) _{Lm}\otimes \chi _{\mu q\overline{q}}^{Sq\overline{q}%
}\right) _{JM}^{PC}
\end{equation}

\begin{eqnarray}
\Psi _{JM}^{PC}(\vec{\rho},\vec{\lambda}) &=&\sum \psi _{l_{g}}^{m_{g}}(\vec{%
\rho})\psi _{l_{q\overline{q}}}^{m_{q\overline{q}}}(\vec{\lambda})\mathbf{e}%
_{\mu _{g}}\chi _{\mu q\overline{q}}^{Sq\overline{q}}\text{ \ }\left\langle
l_{g}m_{g}1\mu _{g}\mid j_{g}M_{g}\right\rangle \left\langle l_{q\bar{q}}m_{q%
\bar{q}}j_{g}M_{g}\mid Lm\right\rangle  \notag \\
&&\left\langle LmS_{q\bar{q}}\mu _{q\bar{q}}\mid JM\right\rangle
\end{eqnarray}

The sum runs over the values of $l_{q\bar{q}},\ m_{q\bar{q}},$ $l_{g},$ $%
m_{g},$ $j_{g},$ $M_{g,}$ $L,$ $m,$ $S_{q\overline{q}},$ $\mu _{q\bar{q}}$
and $\mu _{g}$ excluding those not consistent with \textit{P} and \textit{C}.

Here the Jacobi coordinates are introduced: 
\begin{equation*}
\begin{array}{l}
\vec{\rho}=\vec{r}_{\bar{q}}-\vec{r}_{q}; \\ 
\vec{\lambda}=\vec{r}_{g}-\frac{M_{q}\vec{r}_{q}+M_{\bar{q}}\vec{r}_{\bar{q}}%
}{M_{q}+M_{\bar{q}}}.%
\end{array}%
\end{equation*}

The Hamiltonian is constructed, containing a phenomenological potential
which reproduces the QCD characteristics; its expression has the
mathematical ``Coulomb + Linear'' form, and we take into account also the
additional spin effects.

The basic hypothesis is the use of the relativistic Schr\"{o}dinger-type
wave equation: 
\begin{equation}
\left\{ \sum\limits_{i=1}^{N}\sqrt{\vec{p}_{i}^{\text{ }2}+m_{i}^{\text{ }2}}%
+V_{eff}\right\} \text{ }\Psi (\vec{r}_{i})\text{ }=E\text{ }\Psi (\vec{r}%
_{i}).
\end{equation}

In the case of multiparticle systems, another equivalent wave equation is
more convenient$^{[10]}$: 
\begin{equation}
\left\{ \sum\limits_{i=1}^{N}\left( \frac{\vec{p}_{i}^{\text{ }2}}{2M_{i}}+%
\frac{M_{i}}{2}+\frac{m_{i}^{2}}{2M_{i}}\right) +V_{eff}\right\} \text{ }%
\Psi (\vec{r}_{i})\text{ }=E\text{ }\Psi (\vec{r}_{i})\text{ };
\end{equation}%
where $M_{i}$ are some \textquotedblleft dynamical masses\textquotedblright\
satisfying the conditions: 
\begin{equation}
\frac{\partial E}{\partial M_{i}}=0\text{ };
\end{equation}%
$V_{eff}$ is the average over the color space of chromo-spatial potential:

\begin{eqnarray}
V_{eff} &=&\left\langle V\right\rangle _{color}=\left\langle
-\sum\limits_{i<j=1}^{N}\mathbf{F}_{i}\cdot \mathbf{F}_{j}\text{ }%
v(r_{ij})\right\rangle _{color}  \notag \\
&=&\sum\limits_{i<j=1}^{N}\alpha _{ij}v(r_{ij})\text{ };
\end{eqnarray}
where $v(r_{ij})$ is the phenomenological potential term.

We take a potential which has the form: 
\begin{equation}
v(r_{ij})=v_{CL}(r_{ij})+v_{SD}(r_{ij});
\end{equation}

where the QCD-motivated ''Coulomb.+Linear'' term is of the form: 
\begin{equation}
v_{CL}(r_{ij})=-\frac{\alpha _{s}}{r_{ij}}+\sigma \text{ }r_{ij}+c\text{ };
\end{equation}
the $\alpha _{s}$, $\sigma $, and $c$ may be fitted by experimental data or
taked from lattice and Regge fits.

The spin-dependent term can split into Spin-Spin, Spin-Orbit and Tensor
terms : 
\begin{equation}
v_{SD}=v_{SS}+v_{SO}+v_{T}
\end{equation}
\begin{equation}
\left( v_{SS}\right) _{ij}=\frac{8\pi \alpha _{h}}{3M_{i}M_{j}}\frac{\sigma
_{h}^{3}}{\sqrt{\pi ^{3}}}\exp (-\sigma _{h}^{2}\text{ }r_{ij}^{2})\text{ }%
\mathbf{S}_{i}\cdot \mathbf{S}_{j}\text{ };
\end{equation}
\begin{equation}
\left( v_{SO}\right) _{ij}=\frac{\alpha _{S}}{2r_{ij}^{3}}\left( \frac{%
\mathbf{S}_{i}\cdot \mathbf{r}_{ij}\times \mathbf{p}_{i}}{M_{i}^{2}}-\frac{%
\mathbf{S}_{j}\cdot \mathbf{r}_{ij}\times \mathbf{p}_{j}}{M_{j}^{2}}-\frac{2%
\mathbf{S}_{i}\cdot \mathbf{r}_{ij}\times \mathbf{p}_{j}-2\mathbf{S}%
_{j}\cdot \mathbf{r}_{ij}\times \mathbf{p}_{i}}{M_{i}M_{j}}\text{ }\right) ;
\end{equation}
\begin{equation}
\left( v_{SO}\right) _{ij}=\frac{\alpha _{S}}{2r_{ij}^{3}}\left( \frac{%
\mathbf{S}_{i}\cdot \mathbf{r}_{ij}\times \mathbf{p}_{i}}{M_{i}^{2}}-\frac{%
\mathbf{S}_{j}\cdot \mathbf{r}_{ij}\times \mathbf{p}_{j}}{M_{j}^{2}}-\frac{2%
\mathbf{S}_{i}\cdot \mathbf{r}_{ij}\times \mathbf{p}_{j}-2\mathbf{S}%
_{j}\cdot \mathbf{r}_{ij}\times \mathbf{p}_{i}}{M_{i}M_{j}}\text{ }\right) ;
\end{equation}
\begin{equation}
\left( v_{T}\right) _{ij}=\frac{\alpha _{S}}{M_{i}M_{j}r_{ij}^{3}}\left( 3%
\text{ }\mathbf{S}_{i}\cdot \mathbf{\hat{r}}_{ij}\text{ }\mathbf{S}_{j}\cdot 
\mathbf{\hat{r}}_{ij}-\mathbf{S}_{i}\cdot \mathbf{S}_{j}\text{ }\right) ;
\end{equation}

\underline{\emph{The mass of the constituent gluon}}

An important ingredient of the model is the mass of the constituent gluon $%
m_{g}$. As for quarks, this represents a dynamical mass and is $\exp $ected
to add $0.7\sim 1.0$ $GeV$ to the corresponding quarkonia. The authors of
ref. $[17]$ are generating $800$ $MeV$ constituent gluon mass in the context
of Dynamical Quark Model employing BCS vacuum, a value which is consistent
with the ($1600$) glueball candidate. Furthermore, there is evidence for
massive like dispersion relation for the gluon, with mass ranging from $%
700\sim 1000$ $MeV$, both lattice and from Schwinger-Dyson equations.$^{%
\left[ 18\right] }$

\underline{\emph{The hybrid mass evaluation}}

In earlier estimations$^{\left[ 10\right] }$ we have used $m_{g}=800$ $MeV$,
here we adopt a slightly modified value ($850$ $MeV$) to the best fit of our
results to the ones obtained by lattice calculations of $1^{-+}c\overline{c}%
g $ and $b\overline{b}g$ masses.

We have to solve the wave equation relative to the Hamiltonian: 
\begin{equation}
H=\sum\limits_{i=q,\text{ }\bar{q},\text{ }g}\left( \frac{\vec{p}_{i}^{\text{
}2}}{2M_{i}}+\frac{M_{i}}{2}+\frac{m_{i}^{2}}{2M_{i}}\right) +V_{eff\text{ }%
};
\end{equation}
with, for the hybrid meson : 
\begin{equation}
\begin{array}{l}
\alpha _{q\bar{q}}=-\frac{1}{6}; \\ 
\alpha _{\bar{q}g}=\alpha _{qg}=\frac{3}{2}.%
\end{array}%
\end{equation}

The relative Hamiltonian is given by: 
\begin{equation}
H_{R}=\frac{\vec{p}_{\rho }^{2}}{2\mu _{\rho }}+\frac{\vec{p}_{\lambda }^{2}%
}{2\mu _{\lambda }}+V_{eff}(\vec{\rho},\vec{\lambda})+\frac{M_{q}}{2}+\frac{%
m_{q}^{2}}{2M_{q}}+\frac{M_{\bar{q}}}{2}+\frac{m_{\bar{q}}^{2}}{2M_{\bar{q}}}%
+\frac{M_{g}}{2}+\frac{m_{g}^{2}}{2M_{g}};
\end{equation}

with 
\begin{equation}
\begin{array}{l}
\mu _{\rho }=\left( \frac{1}{M_{q}}+\frac{1}{\text{ }M_{\bar{q}}}\right)
^{-1} \\ 
\mu _{\lambda }=\left( \frac{1}{M_{g}}+\frac{1}{M_{q}+M_{\bar{q}}}\right)
^{-1};%
\end{array}%
\end{equation}

The spatial trial wavefunctions are of the Gaussian-type: 
\begin{eqnarray}
\psi _{l_{g}}^{m_{g}}(\vec{\rho})\psi _{l_{q\overline{q}}}^{m_{q\overline{q}%
}}(\vec{\lambda}) &=&C_{\beta }^{l_{q\bar{q}}}C_{\beta }^{l_{g}}\rho ^{l_{q%
\bar{q}}}\lambda ^{l_{g}}\exp \left( -\frac{1}{2}\beta ^{2}\left( \rho
^{2}+\lambda ^{2}\right) \right) \mathbf{Y}_{l_{q\bar{q}}m_{q\bar{q}%
}}(\Omega _{\rho })\mathbf{Y}_{l_{g}m_{g}}(\Omega _{\lambda })  \notag \\
&=&C_{\beta }^{l_{q\bar{q}}l_{g}}\xi ^{l_{q\bar{q}}+l_{g}}\exp \left( -\frac{%
1}{2}\beta ^{2}\xi ^{2}\right) \mathbf{Y}_{K}(\Omega _{5})
\end{eqnarray}

For a $J^{PC}$ state we calculate the energy function: 
\begin{equation}
En(M_{i},\beta _{l_{q\bar{q}}l_{g}})_{J^{PC}}=\frac{\int \Psi _{JM}^{PC}(%
\vec{\rho},\vec{\lambda})^{\ast }H_{R}\Psi _{JM}^{PC}(\vec{\rho},\vec{\lambda%
})\text{ }d\vec{\rho}d\vec{\lambda}}{\int \Psi _{JM}^{PC}(\vec{\rho},\vec{%
\lambda})^{\ast }\Psi _{JM}^{PC}(\vec{\rho},\vec{\lambda})\text{ }d\vec{\rho}%
d\vec{\lambda}}
\end{equation}

and minimize it to respect of the parameters $M_{i},\beta _{l_{q\bar{q}%
}l_{g}}$ to have both masses and parameters of the corresponding
wavefunctions. For more details see the Appendix. Numerical results are
exposed in tables 2-4.

\section{The decay}

The decay process occurs through the quark pair creation mechanism,
annihilating the constituent glue. Original quarks being spectators; more
details can be found in $\left[ 9\right] .$ The hybrid meson decaying in two
standard mesons obeys to the following (model independent) selection rules:

$\cdot $ only QE-hybrid meson decays in two S-standard mesons;

$\cdot $ only GE-hybrid meson decays in one L and one S-standard mesons.

Some earlier decay results are reviewed taking into account a significative
mixing of the QE and GE modes generating from our mass estimations, this is
summarized in tables 5-8.

\section{Results and discussion}

The mass results are grouped in tables 2-4.

We note that the hybrid masses obtained are in good agreement\ with those
obtained by other models. The spin corrections are more important in the
light sector, and do\ not exceed $19\%$ for $\Delta _{SO}$ and $7\%$. for $%
\Delta _{SS}$ and $\Delta _{Tens}$.

The mass of the GE-hybrid is larger than those in the QE-mode. Indeed the
strong force being proportional to the color charge, the exchange of a color
octet does require an important energy.

In the case of the pure GE mode, we have $%
M_{0^{-+}}<M_{1^{--}}<M_{1^{-+}}<M_{0^{--}}$ (table 4). This is more close
to the results of the ''gluon excitation''\ models; this is simply because
such works exclude the non excited glue ($l_{g}=0$) mode (i.e. the QE one)
from its construction of hybrid mesons.

Note that $1^{-+}n\overline{n}g$ is around $1.93$ $GeV$ ($300\sim 500$ more
heavier than $\pi (1600)$ and $\pi (1400)$ candidates), this at least
exclude that the last candidate to be hybrid meson$^{\left[ 5\right] }$.

On the other hand for $J^{P}=1^{-}$ hybrids we find a significant mixing
between QE and GE-modes. Using the $1^{-+}n\overline{n}g$ wavefunctions
obtained, we have estimate decay widths ( assuming $M_{1^{-+}n\overline{n}%
g}\sim 1.6$ $GeV$ , Tables 5A-6A); the channel $b_{1}\pi $ remains dominant,
and the observation of $1^{-+}n\overline{n}g$ decaying in two S-mesons
remains possible ( $\Gamma _{tot}\left[ 1^{-+}\left( 1600\right) \right]
\simeq 60$ $MeV$). However the mass gap imposes some doubts to consider the
candidate $\pi (1600)$ as an hybrid meson. This conclusion is supported ref. 
$\left[ 6\right] .$

In the case for the candidate $1^{-+}$ around $2.0$ $GeV$ observed by AGS
Coll.$^{[19]},$which is more close to our mass estimations, (Tables 5B-6B),
we note that for the ''L+S''\ channel the decay widths are very large. We
can observe a narrow resonance in the ''S+S'' channel.

In the charm sector, the $1^{--}c\overline{c}g$ is estimated to have a mass
around $4.3$ $GeV$ which is consistent with the Y(4260) candidate.
Otherwise, we find that $1^{--}c\overline{c}g$ decays in two (L+S)-mesons
with decay widths sufficiently small to generate observable resonance $%
\left( \Gamma _{tot}^{1^{--}\left( 4.3\right) }\sim 107\text{ }MeV\right) $.
Note that the possible mixing of $1^{--}c\overline{c}g$ and the
corresponding $1^{--}c\overline{c}$ is excluded$^{\left[ 9\right] }.$

Finally, to have an idea on the geometrical configuration of the hybrid
mesons, we calculate the ratio $\frac{\left\langle \rho ^{2}\right\rangle }{%
\left\langle \lambda ^{2}\right\rangle }$ (table 9); the gluelump picture
that consider the hybrid as a point like $q\bar{q}$ pair bound to a
constituent gluon, is not supported by the present results.\\[1cm]

{\Large Acknowledgments}

This work is supported by the Laboratoire de Physique Th\'{e}orique d'Oran.

\begin{center}
{\huge Appendix: The matrix elements}
\end{center}

The calculation of the matrix elements requires the evaluation of many
(cpu-intensive) six-dimensional integrals. Fortunately, the symmetry of the
problem allows us to do this by an analytic way.

\begin{center}
\underline{{\large The spatial matrix elements}}
\end{center}

Using the hyperspherical coordinates

\ \ 

\QTP{Body Math}
$\mathbf{\xi }\equiv \left( \xi ;\mathbf{\Omega }_{5}\right) \equiv \left(
\xi ;\Omega _{\mathbf{\rho }};\Omega _{\mathbf{\lambda }};\alpha \right)
\equiv \left( \xi ;\theta _{\mathbf{\rho }},\varphi _{\mathbf{\rho }};\theta
_{\mathbf{\lambda }},\varphi _{\mathbf{\lambda }};\theta \right) $

\QTP{Body Math}
$\QATOP{\rho =\xi \sin \theta \text{ }}{\lambda =\xi \cos \theta }$ ;\ \ $%
(0\leq \theta \leq \frac{\pi }{2})$ and \ $\xi ^{2}=\rho ^{2}+\lambda ^{2%
\text{ }};\ \ (0\leq \xi \leq \infty ).$

\ \ 

the potential energy will be more symmetric and we can separate angular from
hyperradial variables:

\begin{center}
$V(\vec{\rho},\vec{\lambda})\equiv V_{\left[ K\right] }(\vec{\xi})=\frac{A_{%
\left[ K\right] }\left( \Omega _{5}\right) }{\xi }+B_{\left[ K\right]
}\left( \Omega _{5}\right) $ $\xi +c^{^{\prime \prime }};$
\end{center}

where $\left[ K\right] \equiv \left( l_{\rho },m_{\rho };l_{\lambda
},m_{\lambda }\right) \equiv \left( l_{q\bar{q}},m_{q\bar{q}%
};l_{g},m_{g}\right) .$

Then matrix elements can be calculated analytically as:

\begin{center}
$V_{\left[ K_{1}K_{2}\right] }=\int\limits_{0}^{\infty }d\xi $ $\xi
^{5}\left( \frac{A_{\left[ K_{1}K_{2}\right] }}{\xi }+B_{\left[ K_{1}K_{2}%
\right] }\text{ }\xi +c^{^{\prime \prime }}\right) R_{\left[ K_{1}\right]
}(\xi )R_{\left[ K_{2}\right] }(\xi );$
\end{center}

where

\begin{center}
$R_{\left[ K_{1}\right] }(\xi )=C_{K}^{l_{\rho }l_{\lambda }}\xi ^{K}\exp
\left( -\frac{1}{2}\alpha _{\left[ K\right] }^{2}\text{ }\xi ^{2}\right) ;$
\ \ \ $K=l_{\rho }+l_{\lambda }$
\end{center}

The coefficients $A_{\left[ K_{1}K_{2}\right] }$ and $B_{\left[ K_{1}K_{2}%
\right] }$ reads: 
\begin{equation*}
\QATOP{\frac{1}{\sigma }B_{K}=\frac{1}{\xi }\int d\Omega _{5}\text{ }\left( -%
\frac{1}{6}\rho +\frac{3}{2}\left| \vec{\lambda}+\frac{\vec{\rho}}{2}\right|
+\frac{3}{2}\left| \vec{\lambda}-\frac{\vec{\rho}}{2}\right| \right) \text{ }%
\mathbf{Y}_{\left[ K_{1}\right] }^{\ast }(\Omega _{5})\mathbf{Y}_{\left[
K_{2}\right] }(\Omega _{5}),}{\frac{1}{\alpha _{s}}A_{K}=-\xi \int d\Omega
_{5}\text{ }\left( -\frac{1}{6\rho }+\frac{3}{2}\frac{1}{\left| \vec{\lambda}%
+\frac{\vec{\rho}}{2}\right| }+\frac{3}{2}\frac{1}{\left| \vec{\lambda}-%
\frac{\vec{\rho}}{2}\right| }\right) \text{ }\mathbf{Y}_{\left[ K_{1}\right]
}^{\ast }(\Omega _{5})\mathbf{Y}_{\left[ K_{2}\right] }(\Omega _{5}),\text{
\ \ \ \ }}
\end{equation*}

with 
\begin{eqnarray*}
\mathbf{Y}_{\left[ K\right] }(\Omega _{5}) &\equiv &\psi _{K}^{l_{\rho
}l_{\lambda }}(\theta )\text{ }\mathbf{Y}_{l_{\rho }m_{\rho }}(\Omega _{%
\mathbf{\rho }})\mathbf{Y}_{l_{\lambda }m_{\lambda }}(\Omega _{\mathbf{%
\lambda }}) \\
&=&N_{K}^{l_{\rho }l_{\lambda }}\sin ^{l\rho }\theta \cos ^{l_{\lambda
}}\theta \text{ }\mathbf{Y}_{l_{\rho }m_{\rho }}(\Omega _{\mathbf{\rho }})%
\mathbf{Y}_{l_{\lambda }m_{\lambda }}(\Omega _{\mathbf{\lambda }}); \\
d\Omega _{5} &\equiv &d\Omega _{\rho }d\Omega _{\lambda }\sin ^{2}\theta
\cos ^{2}\theta \text{ }d\theta \text{ ; \ \ \ \ }d\Omega _{i}\equiv \sin
\theta _{i}\text{ }d\theta _{i}\text{ }d\varphi _{i}.
\end{eqnarray*}

We have used the well known expressions: 
\begin{equation*}
\QATOP{\left| \vec{r}_{1}\pm \vec{r}_{2}\right| \text{ \ }%
=\sum\limits_{l=0}^{\infty }\sum\limits_{m=-l}^{l}\text{ }(\mp )^{l}a_{l}%
\text{ }\frac{4\pi }{2l+1}\text{ }\mathbf{Y}_{lm}(\Omega _{1})\mathbf{Y}%
_{lm}^{\ast }(\Omega _{2}),}{\left| \vec{r}_{1}\pm \vec{r}_{2}\right|
^{-1}=\sum\limits_{l=0}^{\infty }\sum\limits_{m=-l}^{l}\text{ }(\mp
)^{l}b_{l}\text{ }\frac{4\pi }{2l+1}\text{ }\mathbf{Y}_{lm}(\Omega _{1})%
\mathbf{Y}_{lm}^{\ast }(\Omega _{2}),}
\end{equation*}

where 
\begin{gather*}
\QATOP{a_{l}=\frac{1}{2l+3}\frac{r_{<}^{l+2}}{r_{>}^{l+1}}-\frac{1}{2l-1}%
\frac{r_{<}^{l}}{r_{>}^{l-1}},}{b_{l}=\frac{r_{<}^{l}}{r_{>}^{l+1}}.\text{ \
\ \ \ \ \ \ \ \ \ \ \ \ \ \ \ \ \ \ \ \ \ }} \\
\QATOP{r_{<}=\min (r_{1},r_{2}),}{r_{>}=\max (r_{1},r_{2}).}
\end{gather*}

with $\vec{r}_{1}=\vec{\lambda}$ and $\vec{r}_{2}=\frac{1}{2}\vec{\rho}.$

We calculate the integrals:

$\left( I_{a}^{\pm }\right) _{\left[ K_{2}K_{1}\right] }\equiv \xi \int
d\Omega $ $\left| \vec{r}_{1}\pm \vec{r}_{2}\right| ^{-1}$ $\mathbf{Y}_{%
\left[ K_{2}\right] }^{\ast }(\Omega _{5})\mathbf{Y}_{\left[ K_{1}\right]
}(\Omega _{5})=\int\limits_{0}^{\frac{\pi }{2}}\sin ^{2}\theta \cos
^{2}\theta $ $d\theta $ $\psi _{K_{2}}^{l_{\rho 2}l_{\lambda 2}}(\theta
)\psi _{K_{1}}^{l_{\rho 1}l_{\lambda 1}}(\theta )\int d\Omega _{\mathbf{\rho 
}}\mathbf{Y}_{l_{\rho 1}m_{\rho 1}}^{\ast }(\Omega _{\mathbf{\rho }})\mathbf{%
Y}_{l_{\rho 2}m_{\rho 2}}(\Omega _{\mathbf{\rho }})\int d\Omega _{\mathbf{%
\lambda }}$ $\mathbf{Y}_{l_{\lambda 1}m_{\lambda 1}}^{\ast }(\Omega _{%
\mathbf{\lambda }})\mathbf{Y}_{l_{\lambda 2}m_{\lambda 2}}(\Omega _{\mathbf{%
\lambda }})$

$\times \sum\limits_{l=0}^{\infty }\sum\limits_{m=-l}^{l}$ $(\mp )^{l}b_{l}$ 
$\frac{4\pi }{2l+1}$ $\mathbf{Y}_{lm}(\Omega _{\mathbf{\lambda }})\mathbf{Y}%
_{lm}^{\ast }(\Omega _{\mathbf{\rho }}).$

\ \ \ 

and we find

\ \ \ 

$\left( I_{a}^{\pm }\right) _{\left[ K_{2}K_{1}\right] }=\delta _{(m\rho
_{1}+m\rho _{2},m\lambda _{1}+m\lambda _{2})}(-1)^{m\rho _{1}+m\rho
_{2}}\sum\limits_{l=l_{\min }}^{l_{\max }}\left( \mp \right) ^{l}\frac{4\pi 
}{2l+1}$ $T_{l}$

$\times \left\langle l_{\rho 1}m_{\rho 1}\left| \mathbf{Y}_{l(-m\rho
_{1}-m\rho _{2})}\right| l_{\rho 2}m\right\rangle \left\langle l_{\lambda
1}m_{\lambda 1}\left| \mathbf{Y}_{l(m\lambda _{1}+m\lambda _{2})}\right|
l_{\lambda 2}m_{\lambda 2}\right\rangle ,$

$\left\langle l_{1}m_{1}\left| \mathbf{Y}_{lm}\right|
l_{2}m_{2}\right\rangle \equiv \int d\Omega $ $\mathbf{Y}_{l_{1}m_{1}}^{\ast
}(\Omega )\mathbf{Y}_{lm}(\Omega )\mathbf{Y}_{l_{2}m_{2}}(\Omega );$

$l_{\min }=\max (|l_{\rho 2}-l_{\rho 1}|,|l_{\lambda 2}-l_{\lambda 1}|),$

$l_{\max }=\min (l_{\rho 2}+l_{\rho 1},l_{\lambda 2}+l_{\lambda 1});$

\ \ \ 

where

\ \ \ 

$T_{l}\equiv \int\limits_{0}^{\arctan (2)}\sin ^{2}\theta \cos ^{2}\theta $ $%
d\theta $ $\psi _{K_{2}}^{l_{\rho 2}l_{\lambda 2}}(\theta )\psi
_{K_{1}}^{l_{\rho 1}l_{\lambda 1}}(\theta )\frac{\sin ^{l}\theta }{2^{l}\cos
^{l+1}\theta }+\int\limits_{\arctan (2)}^{\frac{\pi }{2}}\sin ^{2}\theta
\cos ^{2}\theta $ $d\theta $ $\psi _{K_{2}}^{l_{\rho 2}l_{\lambda 2}}(\theta
)\psi _{K_{1}}^{l_{\rho 1}l_{\lambda 1}}(\theta )\frac{2^{l+1}\cos
^{l}\theta }{\sin ^{l+1}\theta }$

\ \ \ 

Finally we can write 
\begin{equation*}
\frac{1}{-\alpha _{s}}A_{_{\left[ K_{2}K_{1}\right] }}=-\frac{1}{6}%
\int\limits_{0}^{\frac{\pi }{2}}\sin \theta \cos ^{2}\theta \text{ }d\theta 
\text{ }\psi _{K_{2}}^{l_{\rho 2}l_{\lambda 2}}(\theta )\psi
_{K_{1}}^{l_{\rho 1}l_{\lambda 1}}(\theta )+\frac{3}{2}\text{ }%
(I_{a}^{+}+I_{a}^{-}).
\end{equation*}

In the same way we evaluate the $B_{_{\left[ K_{2}K_{1}\right] }}$%
coefficient: 
\begin{equation*}
\frac{1}{\sigma }B_{K}=-\frac{1}{6}\int\limits_{0}^{\frac{\pi }{2}}\sin
^{3}\theta \cos ^{2}\theta \text{ }d\theta \text{ }\left| \psi _{K}^{l_{\rho
}l_{\lambda }}(\theta )\right| ^{2}+\frac{3}{2}\text{ }(I_{b}^{+}+I_{b}^{-}),
\end{equation*}

with 
\begin{subequations}
\begin{eqnarray*}
\left( I_{b}^{\pm }\right) _{\left[ K_{2}K_{1}\right] } &=&\delta _{(m\rho
_{1}+m\rho _{2},m\lambda _{1}+m\lambda _{2})}(-1)^{m\rho _{1}+m\rho
_{2}}\sum\limits_{l=l_{\min }}^{l_{\max }}\left( \mp \right) ^{l}\frac{4\pi 
}{2l+1}\text{ }S_{l} \\
&&\times \left\langle l_{\rho 1}m_{\rho 1}\left| \mathbf{Y}_{l(-m\rho
_{1}-m\rho _{2})}\right| l_{\rho 2}m_{\rho 2}\right\rangle \left\langle
l_{\lambda 1}m_{\lambda 1}\left| \mathbf{Y}_{l(m\lambda _{1}+m\lambda
_{2})}\right| l_{\lambda 2}m_{\lambda 2}\right\rangle ,
\end{eqnarray*}

where 
\end{subequations}
\begin{eqnarray*}
S_{l} &\equiv &\int\limits_{0}^{\arctan (2)}\sin ^{2}\theta \cos ^{3}\theta 
\text{ }d\theta \text{ }\psi _{K_{2}}^{l_{\rho 2}l_{\lambda 2}}(\theta )\psi
_{K_{1}}^{l_{\rho 1}l_{\lambda 1}}(\theta )\left( \frac{1}{2l+3}\frac{\tan
^{l+2}\theta }{2^{l+2}}-\frac{1}{2l-1}\frac{\tan ^{l}\theta }{2^{l}}\right) +
\\
&&\int\limits_{\arctan (2)}^{\frac{\pi }{2}}\sin ^{3}\theta \cos ^{2}\theta 
\text{ }d\theta \text{ }\psi _{K_{2}}^{l_{\rho 2}l_{\lambda 2}}(\theta )\psi
_{K_{1}}^{l_{\rho 1}l_{\lambda 1}}(\theta )\left( \frac{2^{l+1}}{2l+3}\cot
^{l+2}\theta -\frac{2^{l-1}}{2l-1}\cot ^{l}\theta \right)
\end{eqnarray*}

The terms $A_{_{\left[ K_{2}K_{1}\right] }}$ and $B_{_{\left[ K_{2}K_{1}%
\right] }}$ are expressed in terms of Gamma functions $\Gamma \left(
z\right) $, the ''threeJ symbols'' $\left( 
\begin{array}{ccc}
j_{1} & j2 & j3 \\ 
m_{1} & m_{2} & m_{3}%
\end{array}
\right) $and the incomplete betha functions $B_{z}\left( a,b\right) .$

\begin{center}
\underline{{\large The spin-spin parts}}
\end{center}

In the same way we can calculate the spin-spin corrections using the
expression:

\begin{eqnarray*}
\Delta V_{SS} &=&\frac{8\pi \alpha _{h}}{3}\frac{\sigma _{h}^{3}}{\sqrt{\pi
^{3}}}\{-\frac{1}{6}\frac{1}{M_{q}M_{\overline{q}}}\mathbf{J}\text{ }%
\left\langle \overrightarrow{s}_{q}\cdot \overrightarrow{s}_{\overline{q}%
}\right\rangle \\
&&+\frac{3}{2}\frac{1}{M_{q}M_{g}}\mathbf{J}^{+}\text{ }\left\langle 
\overrightarrow{s}_{q}\cdot \overrightarrow{S}_{g}\right\rangle \\
&&+\frac{3}{2}\frac{1}{M_{\overline{q}}M_{g}}\mathbf{J}^{-}\text{ }%
\left\langle \overrightarrow{s}_{\overline{q}}\cdot \overrightarrow{S}%
_{g}\right\rangle \}.
\end{eqnarray*}

where

\begin{eqnarray*}
\mathbf{J} &=&\int \Psi _{JM}^{PC}(\vec{\rho},\vec{\lambda})^{\ast }\exp
(-\sigma _{h}^{2}\text{ }\rho ^{2})\Psi _{JM}^{PC}(\vec{\rho},\vec{\lambda})%
\text{ }d\vec{\rho}d\vec{\lambda}, \\
\mathbf{J}^{+} &=&\int \Psi _{JM}^{PC}(\vec{\rho},\vec{\lambda})^{\ast }\exp
(-\sigma _{h}^{2}\text{ }\left| \vec{\lambda}+x_{+}\vec{\rho}\right|
^{2})\Psi _{JM}^{PC}(\vec{\rho},\vec{\lambda})\text{ }d\vec{\rho}d\vec{%
\lambda}, \\
\mathbf{J}^{-} &=&\int \Psi _{JM}^{PC}(\vec{\rho},\vec{\lambda})^{\ast }\exp
(-\sigma _{h}^{2}\text{ }\left| \vec{\lambda}-x_{-}\vec{\rho}\right|
^{2})\Psi _{JM}^{PC}(\vec{\rho},\vec{\lambda})\text{ }d\vec{\rho}d\vec{%
\lambda}.
\end{eqnarray*}

We use here the developement: 
\begin{eqnarray*}
\exp \left[ \pm \rho \lambda \cos (\overrightarrow{\rho },\overrightarrow{%
\lambda })\right] &=&\sum\limits_{l=0}^{\infty }\left( \mp \right)
^{l}C_{l}(\rho \lambda )\text{ }P_{l}[\cos (\overrightarrow{\rho },%
\overrightarrow{\lambda })] \\
&=&\sum\limits_{l=0}^{\infty }\left( \mp \right) ^{l}C_{l}(\rho \lambda )%
\text{ }\frac{4\pi }{2l+1}\sum\limits_{m=-l}^{+l}\mathbf{Y}_{lm}^{\ast
}(\Omega _{\mathbf{\rho }})\mathbf{Y}_{lm}(\Omega _{\mathbf{\lambda }})
\end{eqnarray*}

with 
\begin{equation*}
C_{l}(\rho \lambda )=\frac{2l+1}{2}\int\limits_{-1}^{+1}d\mu \text{ }\exp
(\rho \lambda \text{ }\mu )\text{ }P_{l}(\mu ).
\end{equation*}

Finaly we arrive to expressions containing a finite sums which are easy to
evaluate.

\begin{center}
\underline{{\large The spin orbit parts}}
\end{center}

The spin orbit matrix elements can be expressed as: 
\begin{eqnarray*}
\Delta V_{SO} &=&-\frac{1}{6}\frac{\alpha _{s}}{2}\left( \overrightarrow{a}%
_{L}\cdot \overrightarrow{\mathbf{L}}+\overrightarrow{a}_{K}\cdot 
\overrightarrow{\mathbf{K}}\right) \\
&&+\frac{3}{2}\frac{\alpha _{s}}{2}\left( \overrightarrow{A^{+}}\cdot 
\overrightarrow{\mathbf{I}^{+}}_{\lambda P\rho }+\overrightarrow{B^{+}}\cdot 
\overrightarrow{\mathbf{I}^{+}}_{\rho P\rho }+\overrightarrow{C^{+}}\cdot 
\overrightarrow{\mathbf{I}^{+}}_{\lambda P\lambda }+\overrightarrow{D^{+}}%
\cdot \overrightarrow{\mathbf{I}^{+}}_{\rho P\lambda }\right) \\
&&+\frac{3}{2}\frac{\alpha _{s}}{2}\left( \overrightarrow{A^{-}}\cdot 
\overrightarrow{\mathbf{I}^{-}}_{\lambda P\rho }+\overrightarrow{B^{-}}\cdot 
\overrightarrow{\mathbf{I}^{-}}_{\rho P\rho }+\overrightarrow{C^{-}}\cdot 
\overrightarrow{\mathbf{I}^{-}}_{\lambda P\lambda }+\overrightarrow{D^{-}}%
\cdot \overrightarrow{\mathbf{I}^{-}}_{\rho P\lambda }\right)
\end{eqnarray*}

with 
\begin{eqnarray*}
\overrightarrow{a_{L}} &=&-\frac{1}{M_{q}}\left( \frac{1}{M_{q}}+\frac{2}{M_{%
\overline{q}}}\right) \left\langle \overrightarrow{s}_{q}\right\rangle
+\left( -\frac{1}{M_{\overline{q}}^{2}}-\frac{2}{M_{q}M_{\overline{q}}}%
\right) \left\langle \overrightarrow{s}_{\overline{q}}\right\rangle , \\
\overrightarrow{a_{K}} &=&\frac{1}{M_{q}+M_{\overline{q}}}\left( \frac{1}{%
M_{q}}\left\langle \overrightarrow{s}_{q}\right\rangle -\frac{1}{M_{%
\overline{q}}}\left\langle \overrightarrow{s}_{\overline{q}}\right\rangle
\right) ; \\
\overrightarrow{A^{+}} &=&-\frac{1}{M_{q}}\left( \frac{1}{M_{q}}\left\langle 
\overrightarrow{s}_{q}\right\rangle +\frac{2}{M_{g}}\left\langle 
\overrightarrow{S_{g}}\right\rangle \right) , \\
\overrightarrow{A^{-}} &=&-\left( \overrightarrow{A^{+}}\text{ with }%
q\rightarrow \overline{q}\text{ and }\overline{q}\rightarrow q\right) ; \\
\overrightarrow{B^{+}} &=&-\frac{1}{M_{q}\left( M_{q}+M_{\overline{q}%
}\right) }\left( \frac{M_{\overline{q}}}{M_{q}}\left\langle \overrightarrow{s%
}_{q}\right\rangle +\frac{2M_{\overline{q}}}{M_{g}}\left\langle 
\overrightarrow{S_{g}}\right\rangle \right) , \\
\overrightarrow{B^{-}} &=&\left( \overrightarrow{B^{+}}\text{ with }%
q\rightarrow \overline{q}\text{ and }\overline{q}\rightarrow q\right) ; \\
\overrightarrow{C^{+}} &=&-\frac{1}{M_{q}}\left( \frac{1}{\left( M_{q}+M_{%
\overline{q}}\right) }+\frac{2}{M_{g}}\right) \left\langle \overrightarrow{s}%
_{q}\right\rangle -\frac{1}{M_{g}} \\
&&\times \left( \frac{1}{M_{g}}+\frac{2}{\left( M_{q}+M_{\overline{q}%
}\right) }\right) \left\langle \overrightarrow{S_{g}}\right\rangle , \\
\overrightarrow{C^{-}} &=&\left( \overrightarrow{C^{+}}\text{ with }%
q\rightarrow \overline{q}\text{ and }\overline{q}\rightarrow q\right) ; \\
\overrightarrow{D^{+}} &=&-\frac{M_{\overline{q}}}{M_{q}\left( M_{q}+M_{%
\overline{q}}\right) }\left( \frac{1}{\left( M_{q}+M_{\overline{q}}\right) }+%
\frac{2}{M_{g}}\right) \left\langle \overrightarrow{s}_{q}\right\rangle -%
\frac{M_{\overline{q}}}{M_{g}\left( M_{q}+M_{\overline{q}}\right) } \\
&&\times \left( \frac{1}{M_{g}}+\frac{2}{\left( M_{q}+M_{\overline{q}%
}\right) }\right) \left\langle \overrightarrow{S_{g}}\right\rangle , \\
\overrightarrow{D^{-}} &=&-\left( \overrightarrow{D^{+}}\text{ with }%
q\rightarrow \overline{q}\text{ and }\overline{q}\rightarrow q\right) .
\end{eqnarray*}
We have to evaluate the following integrals: 
\begin{eqnarray*}
\overrightarrow{\mathbf{L}} &\equiv &\int \Psi _{JM}^{PC}(\vec{\rho},\vec{%
\lambda})^{\ast }\left( \frac{\vec{\rho}\times \overrightarrow{p}_{\rho }}{%
\rho ^{3}}\right) \Psi _{JM}^{PC}(\vec{\rho},\vec{\lambda})\text{ }d\vec{\rho%
}d\vec{\lambda} \\
\overrightarrow{\mathbf{K}} &\equiv &\int \Psi _{JM}^{PC}(\vec{\rho},\vec{%
\lambda})^{\ast }\left( \frac{\vec{\rho}\times \overrightarrow{p}_{\lambda }%
}{\rho ^{3}}\right) \Psi _{JM}^{PC}(\vec{\rho},\vec{\lambda})\text{ }d\vec{%
\rho}d\vec{\lambda} \\
\overrightarrow{\mathbf{I}^{\pm }}_{iP_{j}} &\equiv &\int \Psi _{JM}^{PC}(%
\vec{\rho},\vec{\lambda})^{\ast }\frac{1}{\left| \vec{\lambda}\pm x_{\pm }%
\vec{\rho}\right| }\overrightarrow{r_{i}}\times \overrightarrow{p}%
_{r_{j}}\Psi _{JM}^{PC}(\vec{\rho},\vec{\lambda})\text{ }d\vec{\rho}d\vec{%
\lambda} \\
\overrightarrow{r}_{i} &=&\vec{\rho}\text{ or }\vec{\lambda}.
\end{eqnarray*}

As an example we calculate $\overrightarrow{I^{\pm }}_{\rho P_{\lambda }}$: 
\begin{equation*}
\overrightarrow{\mathbf{I}^{\pm }}_{\rho P_{\lambda }}\equiv \int \Psi
_{JM}^{PC}(\vec{\rho},\vec{\lambda})^{\ast }\frac{1}{\left| \vec{\lambda}\pm
x_{\pm }\vec{\rho}\right| }\overrightarrow{\rho }\times \left( -%
\overrightarrow{\nabla }_{\lambda }\right) \Psi _{JM}^{PC}(\vec{\rho},\vec{%
\lambda})\text{ }d\vec{\rho}d\vec{\lambda}
\end{equation*}

The particular form of the trial wavefunction leads to expressions: 
\begin{eqnarray*}
\left( \overrightarrow{\mathbf{I}^{\pm }}_{\rho P_{\lambda }}\right) _{k} &=&%
\text{\textbf{\.{I}}}\left[ \alpha _{2}^{2}\left( \mathbf{I}_{\rho
_{i}\lambda _{j}}^{\pm }-\mathbf{I}_{\rho _{j}\lambda _{i}}^{\pm }\right)
+C_{\alpha _{2}}^{l_{\lambda }}\left( -G_{j}^{l_{\lambda }m_{\lambda }}%
\mathbf{J}_{\rho _{i}}^{\pm }+G_{i}^{l_{\lambda }m_{\lambda }}\mathbf{J}%
_{\rho _{j}}^{\pm }\right) \right] ; \\
i,j,k &=&x,y,z\text{ }+circular\text{ }permutation.;\text{with:} \\
\mathbf{I}_{\rho _{i}\lambda _{j}}^{\pm } &=&\mathbf{I}_{\lambda _{j}\rho
_{i}}^{\pm }=\int \Psi _{JM}^{PC}(\vec{\rho},\vec{\lambda})^{\ast }\frac{%
\rho _{i}\lambda _{j}}{\left| \vec{\lambda}\pm x_{\pm }\vec{\rho}\right| ^{3}%
}\Psi _{JM}^{PC}(\vec{\rho},\vec{\lambda})\text{ }d\vec{\rho}d\vec{\lambda},
\\
\mathbf{J}_{\rho _{j}}^{\pm } &=&\int \Psi _{JM}^{PC}(\vec{\rho},\vec{\lambda%
})^{\ast }\frac{\rho _{i}}{\left| \vec{\lambda}\pm x_{\pm }\vec{\rho}\right|
^{3}}\exp \left( -\frac{1}{2}\alpha _{\left[ K_{2}\right] }^{2}\lambda
^{2}\right) \Psi _{JM}^{PC}(\vec{\rho},\vec{\lambda})\text{ }d\vec{\rho}d%
\vec{\lambda}, \\
\mathbf{J}_{\lambda _{j}}^{\pm } &=&\int \Psi _{JM}^{PC}(\vec{\rho},\vec{%
\lambda})^{\ast }\frac{\lambda _{i}}{\left| \vec{\lambda}\pm x_{\pm }\vec{%
\rho}\right| ^{3}}\exp \left( -\frac{1}{2}\alpha _{\left[ K_{2}\right]
}^{2}\rho ^{2}\right) \Psi _{JM}^{PC}(\vec{\rho},\vec{\lambda})\text{ }d\vec{%
\rho}d\vec{\lambda}.
\end{eqnarray*}

These integrals can be calculated using the development: 
\begin{equation*}
\sqrt{1+x^{2}\pm 2x\mu }^{(-3)}=\frac{1}{1-x^{2}}\sum\limits_{l=0}^{\infty
}\left( \mp \right) ^{l}\left( 2l+1\right) x^{l}P_{l}\left( \mu \right) ;%
\text{ \ }x\neq \pm 1
\end{equation*}
then we can write: 
\begin{equation*}
\xi ^{3}\left| \vec{\lambda}\pm x\vec{\rho}\right| ^{\left( -3\right) }=%
\frac{t_{\theta }}{1-x^{2}}\sum\limits_{l=0}^{\infty }\left( \mp \right)
^{l}x^{l}4\pi \sum\limits_{l=-m}^{m}\mathbf{Y}_{lm}^{\ast }(\Omega _{\mathbf{%
\rho }})\mathbf{Y}_{lm}(\Omega _{\mathbf{\lambda }}),
\end{equation*}

If we have the same flavors, then $M_{q}=M_{\overline{q}}$ and 
\begin{equation*}
\begin{array}{ccc}
& 0\prec \theta \leq \arctan 2 & \arctan 2\prec \theta \prec \frac{\pi }{2}
\\ 
t_{\theta } & \cos ^{-3}\theta & 2^{3}\sin ^{-3}\theta \\ 
x & \frac{1}{2}\tan \theta & \frac{2}{\tan \theta }%
\end{array}%
\end{equation*}

In the end we have

$\mathbf{I}_{\rho _{i}\lambda _{j}}^{\pm }=\mathbf{I}_{\lambda _{j}\rho
_{i}}^{\pm }=C_{\alpha _{1}}^{l_{\rho _{1}}l_{\lambda _{1}}}C_{\alpha
_{2}}^{l_{\rho _{2}}l_{\lambda _{2}}}\left( \int\limits_{0}^{\infty }d\xi 
\text{ }\xi ^{4+K_{1}+K_{2}}\exp (-\frac{\alpha _{1}^{2}+\alpha _{2}^{2}}{2}%
\xi ^{2})\right) \sum\limits_{l=l_{\min }}^{l_{\max }}\left( \mp \right)
^{l} $ $\int\limits_{0}^{\frac{\pi }{2}}d\theta $ $\sin ^{3}\theta $ $\cos
^{3}\theta $ $\frac{t_{\theta }}{1-x^{2}}x^{l}4\pi
\sum\limits_{m=-l}^{l}\left( -1\right) ^{m}\left\langle l_{\rho 1}m_{\rho
1}\left| \mathbf{n}_{i}\mathbf{Y}_{lm}\right| l_{\rho 2}m_{\rho
2}\right\rangle \left\langle l_{\rho 1}m_{\rho 1}\left| \mathbf{n}_{j}%
\mathbf{Y}_{lm}\right| l_{\rho 2}m_{\rho 2}\right\rangle ;$

where $\mathbf{n}_{i}=\frac{\mathbf{r}_{i}}{r}.$

\begin{center}
\underline{{\large The tensor part}}
\end{center}

The tensor corrections are calculated from the formula:

\begin{eqnarray*}
\Delta V_{T} &=&-\frac{1}{6}\alpha _{s}\frac{1}{M_{q}M_{\overline{q}}}\left(
\sum\limits_{i,j=x,y,z}3\text{ }\left\langle s_{qi}\text{ }s_{\overline{q}%
j}\right\rangle \text{ }\mathbf{T}_{ij}-\mathbf{T}\text{ }\left\langle 
\overrightarrow{s}_{q}\cdot \overrightarrow{s}_{\overline{q}}\right\rangle
\right) \\
&&+\frac{3}{2}\alpha _{s}\frac{1}{M_{q}M_{g}}\left( \sum\limits_{i,j=x,y,z}3%
\text{ }\left\langle s_{qi}\text{ }S_{gj}\right\rangle \text{ }\mathbf{T}%
_{ij}^{+}-\mathbf{T}^{+}\text{ }\left\langle \overrightarrow{s}_{q}\cdot 
\overrightarrow{S}_{g}\right\rangle \right) \\
&&+\frac{3}{2}\alpha _{s}\frac{1}{M_{\overline{q}}M_{g}}\left(
\sum\limits_{i,j=x,y,z}3\text{ }\left\langle s_{\overline{q}i}\text{ }%
S_{gj}\right\rangle \text{ }\mathbf{T}_{ij}^{-}-\mathbf{T}^{-}\text{ }%
\left\langle \overrightarrow{s}_{q}\cdot \overrightarrow{S}_{g}\right\rangle
\right)
\end{eqnarray*}
with the definitions 
\begin{eqnarray*}
\mathbf{T}_{ij} &=&\int \Psi _{JM}^{PC}(\vec{\rho},\vec{\lambda})^{\ast }%
\frac{\left( \mathbf{n}_{\rho }\right) _{i}\left( \mathbf{n}_{\rho }\right)
_{j}}{\rho ^{3}}\Psi _{JM}^{PC}(\vec{\rho},\vec{\lambda})\text{ }d\vec{\rho}d%
\vec{\lambda}, \\
\mathbf{T} &=& \\
\mathbf{T}_{ij}^{\pm } &=&\int \Psi _{JM}^{PC}(\vec{\rho},\vec{\lambda}%
)^{\ast }\frac{\left( \mathbf{n}_{\pm }\right) _{i}\left( \mathbf{n}_{\pm
}\right) _{j}}{\left| \vec{\lambda}\pm x_{\pm }\vec{\rho}\right| ^{3}}\Psi
_{JM}^{PC}(\vec{\rho},\vec{\lambda})\text{ }d\vec{\rho}d\vec{\lambda}, \\
\mathbf{T}^{\pm } &=&\int \Psi _{JM}^{PC}(\vec{\rho},\vec{\lambda})^{\ast }%
\frac{1}{\left| \vec{\lambda}\pm x_{\pm }\vec{\rho}\right| ^{3}}\Psi
_{JM}^{PC}(\vec{\rho},\vec{\lambda})\text{ }d\vec{\rho}d\vec{\lambda}, \\
\mathbf{n}_{\pm } &=&\frac{\vec{\lambda}\pm x_{\pm }\vec{\rho}}{\left| \vec{%
\lambda}\pm x_{\pm }\vec{\rho}\right| }.
\end{eqnarray*}

To evaluate these integrals we need to exploit the development:

\begin{equation*}
\xi ^{5}\left| \vec{\lambda}\pm x\vec{\rho}\right| ^{\left( -5\right) }=%
\frac{1}{3\left( 1-x^{2}\right) ^{3}}\sum\limits_{l=0}^{\infty }\left( \mp
\right) ^{l}x^{l}\left[ 2l+3-\left( 2l-1\right) x^{2}\right] \text{ }4\pi
\sum\limits_{l=-m}^{m}\mathbf{Y}_{lm}^{\ast }(\Omega _{\mathbf{\rho }})%
\mathbf{Y}_{lm}(\Omega _{\mathbf{\lambda }}),
\end{equation*}

\newpage

\begin{equation*}
\underset{\text{\textit{Table 1: }}l_{q\bar{q}}\text{ \textit{and } }l_{g}%
\text{\textit{\ Parity.}}}{%
\begin{tabular}{|l|l|l|l|l|}
\hline
\textit{P} & \textit{C} & $S_{q\bar{q}}$ & $l_{q\bar{q}}$ & $l_{g}$ \\ \hline
- & - & 
\begin{tabular}{l}
0 \\ 
1%
\end{tabular}
& 
\begin{tabular}{l}
\textit{even} \\ 
\textit{odd}%
\end{tabular}
& 
\begin{tabular}{l}
\textit{odd} \\ 
\textit{even}%
\end{tabular}
\\ \hline
- & + & 
\begin{tabular}{l}
0 \\ 
1%
\end{tabular}
& 
\begin{tabular}{l}
\textit{odd} \\ 
\textit{even}%
\end{tabular}
& 
\begin{tabular}{l}
\textit{even} \\ 
\textit{odd}%
\end{tabular}
\\ \hline
+ & - & 
\begin{tabular}{l}
0 \\ 
1%
\end{tabular}
& 
\begin{tabular}{l}
\textit{even} \\ 
\textit{odd}%
\end{tabular}
& 
\begin{tabular}{l}
\textit{even} \\ 
\textit{odd}%
\end{tabular}
\\ \hline
+ & + & 
\begin{tabular}{l}
0 \\ 
1%
\end{tabular}
& 
\begin{tabular}{l}
\textit{odd} \\ 
\textit{even}%
\end{tabular}
& 
\begin{tabular}{l}
\textit{odd} \\ 
\textit{even}%
\end{tabular}
\\ \hline
\end{tabular}%
\ }
\end{equation*}%
$\medskip $

\begin{equation*}
\underset{\text{\textit{Table 2: Mixed QE-GE hybrid mesons masses with spin
corrections (in GeV)}}}{^{ 
\begin{tabular}[t]{|c|c|c|c|c|}
\hline
& $1^{--}$ & $0^{--}$ & $0^{-+}$ & $1^{-+}$ \\ \hline
$n\overline{n}g$ & 1.40 & 1.58 & 1.73 & 1.93 \\ \hline
$s\overline{s}g$ & 1.66 & 1.87 & 2.02 & 2.21 \\ \hline
$c\overline{c}g$ & 4.27 & 4.35 & 4.38 & 4.48 \\ \hline
$b\overline{b}g$ & 10.50 & 10.66 & 10.68 & 10.80 \\ \hline
\end{tabular}
}}
\end{equation*}

\begin{equation*}
\underset{\text{\textit{Table 3: Pure QE and GE-hybrid masses (in GeV)}}}{%
\begin{tabular}[t]{|c|c|c|}
\hline
& $QE$ & $GE$ \\ \hline
$n\overline{n}g$ & 1.31 & 1.70 \\ \hline
$s\overline{s}g$ & 1.57 & 2.00 \\ \hline
$c\overline{c}g$ & 4.09 & 4.45 \\ \hline
$b\overline{b}g$ & 10.34 & 10.81 \\ \hline
\end{tabular}%
\ }
\end{equation*}

\begin{equation*}
\underset{\text{\textit{Table 4: Pure GE-light hybrid masses with spin
corrections (in GeV)}}}{%
\begin{tabular}{|c|c|c|c|c|}
\hline
& $0^{-+}$ & $1^{--}$ & $1^{-+}$ & $0^{--}$ \\ \hline
$n\overline{n}g$ & 1.83 & 1.90 & 2.04 & 2.06 \\ \hline
$s\overline{s}g$ & 2.13 & 2.21 & 2.33 & 2.38 \\ \hline
\end{tabular}%
\ }
\end{equation*}%
\newpage

\begin{equation*}
\underset{\text{\textit{Table 5A: Decay widths of the (M=1.6) hybrid in
(S+S)-standard mesons (in }}\alpha _{s}\text{\textit{MeV).}}}{ 
\begin{tabular}{|c|c|}
\hline
$\Gamma _{\rho \omega }$ & $1.5$ \\ \hline
$\Gamma _{\rho \pi }$ & $53$ \\ \hline
$\Gamma _{K^{\ast }K}$ & $7.0$ \\ \hline
$\Gamma _{tot}\left[ 1^{-+}\left( 1600\right) \right] $ & $61$ \\ \hline
\end{tabular}
}
\end{equation*}
\begin{equation*}
\underset{\text{\textit{Table 5B: Decay widths of the (M=2.0) hybrid in
(S+S)-standard mesons (in }}\alpha _{s}\text{\textit{MeV).}}}{ 
\begin{tabular}{|l|l|}
\hline
$\Gamma _{\rho \omega }$ & $31$ \\ \hline
$\Gamma _{\rho \pi }$ & $100$ \\ \hline
$\Gamma _{\rho (1450)\pi }$ & $18$ \\ \hline
$\Gamma _{K^{\ast }K}$ & $32$ \\ \hline
$\Gamma _{K^{\ast }(1410)K}$ & $0.3$ \\ \hline
$\Gamma _{tot}\left[ 1^{-+}\left( 2000\right) \right] $ & $181$ \\ \hline
\end{tabular}
}
\end{equation*}

\begin{equation*}
\underset{\text{\textit{Table 6A: Decay widths of the (M=1.6) hybrid in
(L+S)-standard mesons (in }}\alpha _{s}\text{\textit{MeV).}}}{ 
\begin{tabular}{|c|c|c|c|}
\hline
$L$ & $0$ & $1$ & $2$ \\ \hline
$\Gamma _{b_{1}^{0}\pi ^{-}}$ & $145$ & $433$ & $722$ \\ \hline
$\Gamma _{b_{1}^{-}\pi ^{0}}$ & $145$ & $434.7$ & $724.5$ \\ \hline
$\Gamma _{f_{1}^{0}(1285)\pi ^{-}}$ & $117$ & $88.2$ & $147.4$ \\ \hline
$\Gamma _{f_{1}^{0}(1420)\pi ^{-}}$ & $39$ & $29$ & $47.9$ \\ \hline
$\Gamma _{tot}\left[ 1^{-+}\left( 1600\right) \right] $ & $446$ & $563$ & $%
1642$ \\ \hline
\end{tabular}
}
\end{equation*}

\begin{equation*}
\underset{\text{\textit{Table 6B: Some partial decay widths of the (M=2.0)
hybrid in (L+S)-standard mesons (in }}\alpha _{s}\text{\textit{MeV).}}}{%
\begin{tabular}{|c|c|c|c|}
\hline
$L$ & $0$ & $1$ & $2$ \\ \hline
$\Gamma _{b_{1}^{0}\pi ^{-}}$ & $477$ & $1425$ & $2378$ \\ \hline
$\Gamma _{f_{1}^{0}(1285)\pi ^{-}}$ & $481$ & $361$ & $604$ \\ \hline
$\Gamma _{f_{1}^{0}(1420)\pi ^{-}}$ & $524$ & $340$ & $643$ \\ \hline
\end{tabular}%
\ }
\end{equation*}

\begin{equation*}
\underset{\text{\textit{Table 7: Decay widths of the (M=4.26) hybrid in
(S+S)-standard mesons (in }}\alpha _{s}\text{\textit{MeV).}}}{_{\underset{%
\text{\textit{.}}}{ 
\begin{tabular}{|c|c|c|c|}
\hline
$L$ & $0$ & $1$ & $2$ \\ \hline
$\Gamma _{D^{0}\bar{D}^{0}}$ & $129.5$ & $388.5$ & $647.5$ \\ \hline
$\Gamma _{D^{+}D^{-}}$ & $135.1$ & $406$ & $676.2$ \\ \hline
$\Gamma _{D_{s}^{+}D_{s}^{-}}$ & $142.8$ & $428.4$ & $714$ \\ \hline
$\Gamma _{D^{\ast 0}\overline{D^{0}}}=\Gamma _{D^{\ast 0}\overline{D^{\ast 0}%
}}$ & $0.00$ & $0.00$ & $0.00$ \\ \hline
$\Gamma _{D^{\ast +}D^{\ast -}=}\Gamma _{D^{\ast 0}\overline{D^{\ast 0}}}%
\text{\ } 
\begin{tabular}{l}
$S=0$ \\ 
$S=1$ \\ 
$S=2$%
\end{tabular}
$ & 
\begin{tabular}{l}
$30.8$ \\ 
\multicolumn{1}{c}{$0.00$} \\ 
$49.3$%
\end{tabular}
& 
\begin{tabular}{l}
$92.4$ \\ 
\multicolumn{1}{c}{$0.00$} \\ 
$369.6$%
\end{tabular}
& 
\begin{tabular}{l}
$1.47$ \\ 
\multicolumn{1}{c}{$0.00$} \\ 
$24.5$%
\end{tabular}
\\ \hline
$\Gamma _{tot}\left[ 1^{--}\left( 4.26\right) \right] $ & $982.1$ & $2463.3$
& $2914.1$ \\ \hline
\end{tabular}
}}}
\end{equation*}

\begin{equation*}
\underset{\text{\textit{Table 8: Partial decay widths of the (M=4.3) hybrid
in (L+S)-standard mesons (in }}\alpha _{s}\text{\textit{MeV).}}}{%
\begin{tabular}{|c|c|}
\hline
$\Gamma _{D_{1}(2420)\overline{D^{0}}}=\Gamma _{\overline{D_{1}}(2420)D^{0}}$
$\simeq \Gamma _{D_{1}^{\pm }(2420)D^{\mp }}$ & $\frac{107}{4}$ \\ \hline
\end{tabular}%
\ }
\end{equation*}

\begin{equation*}
\underset{\text{Table 9: \textit{Geometrical configuration of the hybrid
mesons.}}}{ 
\begin{tabular}{|c|c|c|c|}
\hline
$\frac{\left\langle \rho ^{2}\right\rangle }{\left\langle \lambda
^{2}\right\rangle }$ & $1^{-+}$ & $0^{-+}$ & $1^{--}$ \\ \hline
\textit{light sector} & 0.8 & 1.0 & 1.4 \\ \hline
\textit{b-sector} & 0.6 & 0.8 & 1.1 \\ \hline
\end{tabular}
}
\end{equation*}

\end{document}